\documentclass[aps,prc,twocolumn,showpacs]{revtex4}
\usepackage{graphicx}
\begin{document}
\title{Scaling Behavior at High $p_T$ and the $p/\pi$ Ratio}
\author{Rudolph C. Hwa$^1$ and C.\ B.\ Yang$^{1,2}$}
\affiliation{$^1$Institute of Theoretical Science and Department of
Physics\\ University of Oregon, Eugene, OR 97403-5203, USA\\
$^2$Institute of Particle Physics, Hua-Zhong Normal
University, Wuhan 430079, P.\ R.\ China}
\date{January 2003}
\begin{abstract}
We first show that the pions produced at high $p_T$ in heavy-ion
collisions over a wide range of high energies exhibit a scaling
behavior when the distributions are plotted in terms of a scaling
variable. We then use the recombination model to calculate the
scaling quark distribution just before hadronization. From the quark
distribution it is then possible to calculate the proton
distribution at high
$p_T$, also in the framework of the recombination model. The
resultant $p/\pi$ ratio exceeds one in the intermediate $p_T$ region
where data exist, but the scaling result for the proton distribution
is not reliable unless $p_T$ is high enough to be insensitive to the
scale-breaking mass effects.

\end{abstract}

\maketitle

\section{Introduction}

There are three separate and independent aspects about the
hadrons produced at large transverse momentum $(p_T)$ in
heavy-ion collisions at high energies that collectively contribute
to a coherent picture to be addressed in this paper.  One is the
existence of a scaling behavior at large $p_T$ that we have found
by presenting the data in terms of a new variable.  Another is the
issue about the surprisingly large proton-to-pion ratio at
moderate $p_T$ ($\sim$ 2 - 3 GeV/c) discovered by PHENIX
\cite{ts} in central $AuAu$ reactions at $\sqrt{s} =$ 130
and 200 GeV.  The third issue concerns the
hadronization process relevant for the formation of
hadrons at large
$p_T$ and the applicability of the recombination model
\cite{dh}.  It is our goal to show that, in light of the scaling
behavior of the $\pi^0$ produced, the recombination
mechanism naturally gives rise to a $p/\pi$ ratio that
exceeds 1 in the $2 < p_T < 3$ GeV/c range.

Particle production  in heavy-ion collisions at very high energies
is usually described in terms of hydrodynamical flow \cite{hyd},
jet production at high $p_T$ \cite{jet}, thermal statistical model
\cite{the}, or a combination of various hadronization mechanisms
\cite{all}.  In none of the conventional approaches does one
expect protons to be produced at nearly the same rate as the
pions. If all hadrons with $p_T > 2$ GeV/c are regarded as
products of jet fragmentation, then the known fragmentation
functions of quark or gluon jets would suppress proton relative to
pion by the sheer weight of the proton mass.  Such a discrepancy
from the observed data led some to regard the
situation as an anomaly and proposed the gluonic baryon junction
as a mechanism to enhance the proton production rate \cite{vg}.
Their predictions remain to be checked by experiments.

The parton fragmentation functions have been used even at low
$p_T$ in string models where the production of particles in
hadronic collisions is treated as the fragmentation of diquarks, as
done in the dual parton model \cite{dpm}.  There has been a
long-standing dichotomy on whether particle production in the
fragmentation region can better be described by fragmentation
\cite{dpm,lpy} or recombination \cite{dh,hy}.  It is
possible that the two pictures might be unified in a more
comprehensive treatment of hadronization in the future.
Here we extend the recombination model to the central
region at large $p_T$.  It should be recognized that an
essential part of the recombination model is the
determination of the distribution functions of the quarks
and antiquarks that are to recombine.  In the case of
large-$p_T$ hadrons the underlying physics is
undoubtedly hard collisions of partons and the associated
radiation of gluons.  If the parton distributions can be
calculated just before hadronization, then the final step of
recombination can readily be completed.  If those
distributions cannot be determined in pQCD, then the step
between the initiating large-$p_T$ parton and the
resultant hadrons may efficiently be described by a
fragmentation function, determined phenomenologically
from experiments.  Thus in that sense the two approaches,
recombination and fragmentation, are not contradictory,
but complementary.

We state from the outset that no attempt will be made here to
perform a first-principle calculation of the parton distributions
at large $p_T$ before recombination.  However, from the
observed data on pion production in central $AuAu$ collisions at
the Relativistic Heavy-Ion Collider (RHIC), it is possible to work
backwards in the recombination model to determine the quark
(and antiquark) distribution at large $p_T$.  On the basis of the
quark distributions inferred, it is then possible to calculate the
proton distribution in the recombination model.  The basic idea
is that if there is a dense system of quarks and antiquarks
produced in a heavy-ion collision whatever the dynamics
responsible for them may have been (gluons having been
converted to $q\bar{q}$ pairs before hadronization), then the
formation of pions and protons (and whatever else) is prescribed
by the recombination model without any arbitrariness in
normalization and momentum dependence.

One limitation of the recombination model as it stands at
present is that it is formulated in a frame-independent way in
terms of momentum fractions and is therefore inapplicable to a
system where the particle momenta are low and the mass effects
are large.  The physics of recombination is still valid at low
momentum, but the details of the wave functions of the
constituent quarks become important; they have not been built into
the recombination function that takes the simplest form in the
infinite momentum frame.  Thus our calculation of particles
produced at midrapidity is not reliable when $p_T$ is of the
order of the masses of the hadrons under consideration.  For
protons we can trust the results only for $p_T > 3$ GeV/c.  For
pions the lower limit of validity can be pushed much lower.

Since our approach makes crucial use of the experimental data on
the pion spectrum as the input, it is essential to relate the
spectra determined at different energies to an invariant
distribution so that the scale-invariant recombination model can
be applied.  To discover the existence of an invariant distribution
with no theoretical prejudices is a problem worthy in its own
right.  Fortunately, that turns out to be possible.  The analysis for
that part of the study will be presented below first to emphasize
its independence from the theoretical modeling of hadronization. It
should be mentioned that the scaling of transverse mass spectra has
been investigated recently \cite{sb}. The emphasis there has been on
the dependences on the particle species and centrality for $m_T<3.8$
GeV, while our focus is on the dependence on energy ($17<\sqrt s<200
$ GeV) for $p_T<8$ GeV/c. Thus the two studies are complementary to
each other.

\section{A Universal Scaling Distribution}

The preliminary data of the $p_T$ distributions of $\pi^0$ produced
at RHIC at
$\sqrt{s} = 130$ and 200 GeV were shown by the PHENIX
Collaboration at Quark Matter 2002 \cite{ddl} for central
$AuAu$ collisions together with the WA98 data for $PbPb$
collisions at $\sqrt{s} = 17$ GeV \cite{rey}.  They show that the
level of the tail at large $p_T$ rises , as
$\sqrt{s}$ is increased.  We want to consider the possibility that
the three sets of data points can be combined to form a universal
curve.

The $\pi^0$ inclusive distributions at midrapidity are
integrated over $\eta$ for a range of $\Delta \eta = 1$ so that the
data points are given for the following quantity \cite{ddl}:
\begin{eqnarray}
f(p_T, s) = {1 \over 2 \pi p_T}{dN \over dp_T}=
\int_{\Delta\eta} d\eta \left(2 \pi p_TN_{evt} \right)^{-1}
{d^2N_{\pi^0}
\over  d p_T d\eta}  .
\label{1}
\end{eqnarray}
In comparing the PHENIX data with those of WA98 one should
recognize that in addition to the difference in the colliding
nuclei there is a slight mismatch in centrality (top 10\% for
PHENIX and top 12.7\% for WA98) \cite{we}.

To unify the three data sets it is natural to first consider a
momentum fraction variable similar to $x_F$  in longitudinal
momentum.  However, so much momenta are taken by the other
particles outside the $\Delta \eta = 1$ range, it is unwise to also
use $\sqrt{s}/2$ as the scale to calculate the transverse
momentum fraction.  We assume that for every $\sqrt{s}$ there is
a relevant scale $K$ to describe the $p_T$ behavior relative to
that scale.  Let us define
\begin{eqnarray}
z = p_T/K,
\label{2}
\end{eqnarray}
and transform $f(p_T, s)$ to a new function $\Phi (z,K)$, where
\begin{eqnarray}
\Phi (z,K) = K^2 f(p_T, s) = {1  \over  2 \pi z} {dN  \over  dz} .
\label{3}
\end{eqnarray}
We adjust $K$ for each $s$ and check whether all three data sets
coalesce into one universal dependence on $z$, which we would
simply label as $\Phi(z)$, if it is possible.

In Fig.\ 1 we show $\Phi(z)$, where the three symbols represent
the three data sets for the three energies.  Evidently, the
universality exists and is striking.  While this behavior needs to
be confirmed by more data, and the theoretical implication
remains to be explored, the existence of this scaling behavior is a
significant phenomenological property of the $p_T$
distributions that suggests some underlying simplicity.  It is like
the KNO scaling of the multiplicity distributions $P(n, s)$ in $pp$
collisions, where for $\sqrt{s} < 200$ GeV they can be expressed
by one universal scaling function $\psi (z)$, with $z =
n/\left<n\right>$
\cite{kno,lat}.

The values of $K$ that are used for the plot in Fig.\ 1 are in
units of GeV:
$K = 1 \, (200)$,  $0.9 \, (130)$ and $0.717 \, (17)$, the quantities in
the parentheses being the values of $\sqrt{s}$.  The $\sqrt{s}$
dependence of $K$ forms nearly a straight line, as shown  in Fig.\
2.  Since the high and low energy data differ both in colliding
nuclei and in centrality, one does not expect strict regularity in
how $K$ depends on $\sqrt{s}$.  Nevertheless, an approximate
linear dependence is a simple behavior expected on dimensional
grounds.  The straight line in Fig.\ 2 corresponds to the best fit
\begin{eqnarray}
K(s) = 0.69 + 1.55 \times 10^{-3} \sqrt{s} ,
\label{4}
\end{eqnarray}
where $\sqrt{s} $ is in units of GeV.  It should be recognized that
the normalization of $K(s)$ is arbitrary; it is chosen to be $1$ at
$\sqrt{s} = 200$ GeV for simplicity.  If it is normalized to some
other value at that point, the linear behavior in Fig.\ 2 is
unchanged, only the scale of the vertical axis is shifted
accordingly.  The scaling property in Fig.\ 1 is also unchanged,
the only modifications being the scales of the horizontal and
vertical axes.  Thus the absolute magnitude of the dimensionless
variable $z$ has no significance.

If the $z$ dependence of $\Phi(z)$ in Fig.\ 1 were strictly linear,
so that it is a power-law dependence
\begin{eqnarray}
\Phi(z) \propto z^{\alpha}  ,
\label{5}
\end{eqnarray}
then there would be no relevant scale in the problem.  The fact
that it is not a straight line implies that there is an intrinsic scale
in the $p_T$ problem, which is hardly surprising.  What is
significant is that while there is no strict scaling in $z$, there is
no explicit dependence on $s$.  That is, at any energy we have
the same universal function $\Phi (z)$, which will be referred to as
the scaling behavior in $s$.  That function can be parametrized
by
\begin{eqnarray}
\Phi (z) = 1500 \left(z^2 + 2 \right)^{-4.9}  ,
\label{6}
\end{eqnarray}
which is represented by the smooth curve in Fig.\ 1.  For large
enough $z$ Eq.\ (\ref{6}) does have the form of the power law
given in Eq.\ (\ref{5}) with $\alpha = 9.8$.  It is a succinct
statement of the universal properties at high $p_T$.  The
departure from Eq.\ (\ref{5}) at small $z$ reflects the physics at
low $p_T$. Since there is no data on $\pi^0$ for $p_T<1$ GeV/c,
the extrapolation of $\Phi(z)$ to $z<1$ is not reliable. However,
there is a more accurate determination of $\Phi(z)$ that includes
the low $z$ region when the charge $\pi^+$ data are considered; it
is given in \cite{hy4}, and is not needed here.

\begin{figure}[tbph]
\includegraphics[width=0.55\textwidth]{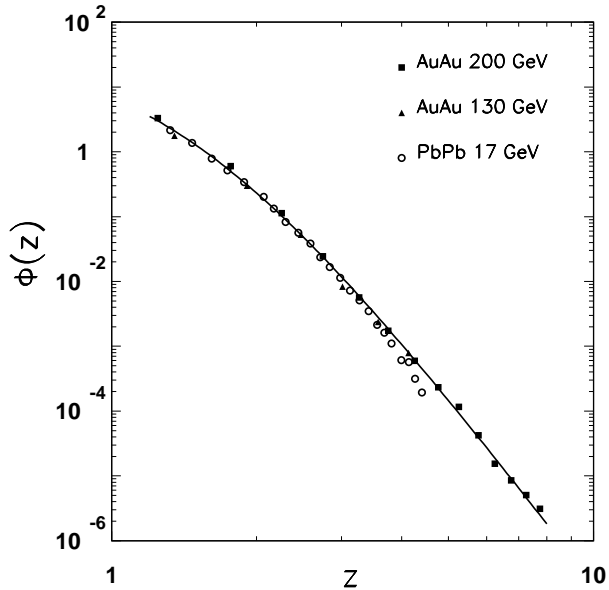}
\caption{Scaled transverse momentum distribution of
produced $\pi^0$. Data are from Ref.\ \cite{ddl,rey}. The solid line
is a fit of the data by Eq.\ (\ref{6}).}
\end{figure}

Note that there is no fixed scale in $p_T$ that separates the high-
and low-$p_T$ physics.  Equation (\ref{6}) gives a smooth
transition from one to the other in the variable $z$, thus
implying different ranges of values of the transition $p_T$ at
different $s$.

While Eq.\ (\ref{6}) gives a good parametrization of the scaling
function $\Phi (z)$ throughout the whole range of $z$, one
notices, however, that the WA98 data at 17 GeV shows a slight
departure from $\Phi (z)$ at the high $z$ end of that data set.  It
should be recognized that those data points have $p_T >3$
GeV/c, which represents a huge fraction of the available energy at
$\sqrt{s} = 17$ GeV.  In fact, one expects the violation of
universality to be more severe at higher $z$ at that $\sqrt{s}$,
since energy conservation would suppress the inclusive cross
section at higher
$p_T$.  What is amazing is that most of the WA98 data points are
well described by $\Phi (z)$, even though the corresponding
$p_T$ values take up a much larger fraction of the available
energy than the other data points from RHIC.  It demonstrates the
significance of the variable $z$ in revealing the scaling property.
\begin{figure}[tbph]
\includegraphics[width=0.55\textwidth]{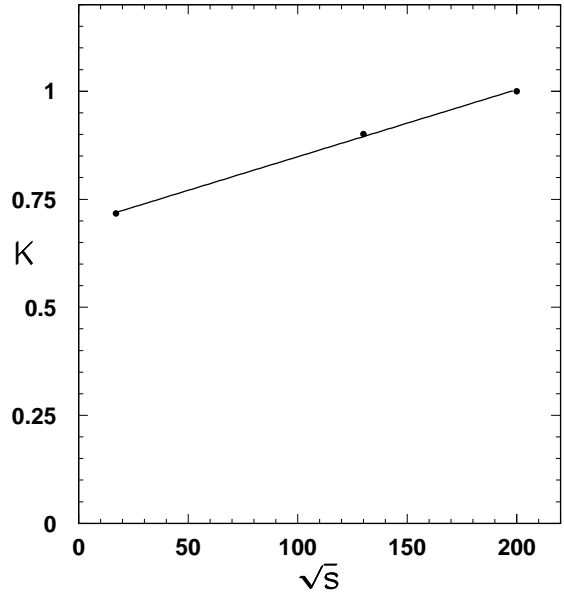}
\caption{The dependence of $K(s)$ on $\sqrt s$. The line
is a linear fit.}
\end{figure}

\section{Pion and Quark Distributions in the Recombination
Model}

Having found a scaling distribution for the produced $\pi^0$
independent of $s$, we now consider the hadronization process in
the recombination model in search for an origin of such a scaling
behavior.  In previous investigations the recombination model
has been applied only to the fragmentation region where the
longitudinal momenta are large and the transverse momenta are
either held fixed at low $p_T$ or integrated over
\cite{dh,hy,hy2}.  We now consider the creation of
pions in the central region of $AA$ collisions and study the $p_T$
dependence.  Unlike the former case where the longitudinal
momentum fractions of the partons are essentially known (from
the structure functions), the $p_T$ distributions of the partons in
the latter case are essentially unknown.  Indeed, it is the aim of
this section to determine the parton $p_T$ distributions from the
$\pi^0$ distribution found in the previous section.

Let us start by writing down the basic equation for recombination
in the 3-space
\begin{eqnarray}
E{d^3N_{\pi}\over d^3p} = \int {d^3p_1  \over  E_1}{d^3p_2
\over  E_2}\ {\cal F}(\vec{p}_1, \vec{p}_2)\,{\cal
R}_{\pi}(\vec{p}_1,
\vec{p}_2, \vec{p})
\label{7}
\end{eqnarray}
where the left-hand side (LHS) is the inclusive distribution of
pion with energy-momentum $(E, p)$.  ${\cal F}(\vec{p}_1,
\vec{p}_2)$ is the probability of having a quark at
$p^{\mu}_1$ and an antiquark at $p^{\mu}_2$ just before
hadronization.  ${\cal R}_{\pi}(\vec{p}_1,
\vec{p}_2, \vec{p})$ is the invariant distribution,
$E{d^3N_{\pi}^{q\bar{q}} /d^3p}$, of producing a pion at
$p^{\mu}$ given a $q$ at $p^{\mu}_1$ and a $\bar{q}$ at
$p^{\mu}_2$.  Note that ${\cal R}_{\pi}$ has the dimension
(momentum)$^{-2}$, same as the LHS.

Writing the phase-space density in the form
\begin{eqnarray}
{d^3p\over E} = dy \, d\phi \, p_T \, dp_T ,
\label{8}
\end{eqnarray}
we define the inclusive distribution in $p_T$, averaged over $y$
and $\phi$,
\begin{eqnarray}
{d^3N_{\pi} \over p_T \,dp_T} = {1   \over  \Delta y}
\int_{\Delta y } dy \ {1 \over 2 \pi} \int^{2\pi}_0 \, d\phi  \
E{d^3N_{\pi}\over d^3p}  ,
\label{9}
\end{eqnarray}
where $\Delta y$ is limited to one unit of rapidity in the central
region.  Our focus will be on the $p_T$ distribution at high
$p_T$.  For the recombination distribution ${\cal
R}_{\pi}(\vec{p}_1, \vec{p}_2, \vec{p})$ we need
only consider the partons in the same transverse plane that
contains
$\vec{p}$, since at high
$p_T$ the partons with different $y_i$ are not likely to
recombine.  Indeed, we assume not only $y_1 = y_2 = y$, but
also $\phi _1 = \phi _2 = \phi$ so that the partons and the pion
are all colinear, and the kinematics can be reduced to that of a
1-dimensional problem.  As in the usual parton model, the parton
momentum fractions in the hadron can vary between 0 and 1,
but the deviation in the momentum components of the partons
transverse to the hadron $\vec{p}$ must be severely limited
because of the limited transverse size of the hadron.  Thus we write
\begin{eqnarray}
{\cal R}_{\pi}(\vec{p}_1, \vec{p}_2,\vec{p}) = {\cal
R}_{\pi} ^0 \  \delta \left(y_1 - y_2\right) \delta \left(\phi _1 -
\phi _2\right)\nonumber\\ \delta \left({y_1 + y_2 \over 2} -y
\right) \delta^2 (\vec{p}_{1_T} + \vec{p}_{2_T} -
\vec{p}_T ) ,
\label{10}
\end{eqnarray}
where ${\cal R}_{\pi}^0 $ is dimensionless, since
$\delta^2 (\vec{p}_{1_T} + \vec{p}_{2_T} -
\vec{p}_T)$ has the dimension of ${\cal
R}_{\pi} (\vec{p}_1, \vec{p}_2, \vec{p})$.  If this
$\delta$-function is further written in the colinear form due to
the $\delta(\phi _1 - \phi _2)$ in Eq.\ (\ref{10})
\begin{eqnarray}
\delta^2 (\vec{p}_{1_T} + \vec{p}_{2_T} -
\vec{p}_T) &=& \delta \left({\phi _1 +
\phi _2 \over 2} - \phi\right) {1 \over p_T} \,
\nonumber\\  &&\delta\left(p_{1_T} + p_{2_T} -
p_T\right) ,
\label{11}
\end{eqnarray}
then Eq.\ (\ref{7}) can be reduced to the 1D form
\begin{eqnarray}
{dN_{\pi}\over p_T dp_T} &=& \int
dp_{1_T} dp_{2_T}p_{1_T}p_{2_T}\ {\cal F}(p_{1_T},
p_{2_T})\,{\cal R}_{\pi} ^0  \, p_T^{\ -2}\nonumber\\ &&\delta\left( {
p_{1_T} + p_{2_T} \over  p_T} -1 \right),
\label{12}
\end{eqnarray}
where ${\cal F}(p_{1_T}, p_{2_T})$ is the $q\bar{q}$
distribution in $p_{i_T}$ averaged over $y$ and $\phi$.

We can reexpress this equation in terms of the scaling variable $z
= p_T/K$, introduced in Eq.\ (\ref{2}), and obtain
\begin{eqnarray}
{dN_{\pi}\over zdz} = \int dz_1 dz_2\,z_1\,z_2\ F(z_1, z_2)\ R_{\pi}
(z_1, z_2, z)
\label{13}
\end{eqnarray}
where
\begin{eqnarray}
F(z_1, z_2) = K^4\ {\cal F}(  p_{1_T} , p_{2_T})
\label{14}
\end{eqnarray}
\begin{eqnarray}
R_{\pi} (z_1, z_2, z) = {\cal R}_{\pi}^0 \ z^{-2}\ \delta\left( { z_1
+ z_2 \over  z}-1
\right)  .
\label{15}
\end{eqnarray}
Since ${\cal F}(  p_{1_T} , p_{2_T} )$ is the parton
density in $p_{1_T} dp_{1_T} p_{2_T} dp_{2_T}$, $F(z_1,
z_2)$ is the corresponding dimensionless density in $z_1
dz_1 z_2 dz_2$.  Equation (\ref{13}) is now our basic formula for
recombination in the scaled transverse-momentum variable.  The
total number of $q$ and $\bar{q}$ is $\int dz_1 dz_2\,z_1z_2
F(z_1, z_2)$, which should be invariant under a change of scale
\begin{eqnarray}
z = \lambda x
\label{16}
\end{eqnarray}
so that
\begin{eqnarray}
x = p_T/K^{\prime}, \qquad \qquad K^{\prime} = \lambda K  .
\label{17}
\end{eqnarray}
The corresponding change on $F(z_1, z_2)$ is that it
becomes
\begin{eqnarray}
F^{\prime} (x_1, x_2 ) = \lambda^4 F(z_1, z_2).
\label{18}
\end{eqnarray}
Thus the normalization of  $F(z_1, z_2)$ is scale
dependent, as it should in view of Eq.\ (\ref{14}).

So far the recombination function $R_{\pi}(z_1, z_2, z)$ is not
fully specified because ${\cal R}_{\pi}^0$ has not been.  In Eq.\
(\ref{15}) the factor $z^{-2}$ is associated with the dimension of
the pion density, and the $\delta$-function with momentum
conservation.  To introduce the pion wave function in terms of the
constituent quarks, we rewrite Eq.\ (\ref{15}) as
\begin{eqnarray}
R_{\pi}(z_1, z_2, z )=R_{\pi}^0\ z^{-2}\ G_{\pi}(\xi_1,\xi _2),
\label{19}
\end{eqnarray}
where $R_{\pi}^0$ is a normalization constant to be determined
and $G_{\pi}(\xi _1,\xi _2)$ is the valon distribution
of the pion \cite{dh,hy}.  Since the recombination of a $q$ and
$\bar{q}$ into a pion is the time-reversed process of
displaying the pion structure, the dependence of
$R_{\pi}(z_1, z_2, z)$ on the pion structure is expected.  During
hadronization the initiating $q$ and $\bar{q}$ dress themselves
and become the valons of the produced hadron without
significant change in their momenta.  The variable $\xi _i$ in
Eq.\ (\ref{19}) denotes the momentum fraction of the $i$th
valon, i.e.,
\begin{eqnarray}
\xi _i = z_i/z  ,
\label{20}
\end{eqnarray}
which is denoted by $y_i$ in the valon model \cite{dh,hy}, a
notation that cannot be repeated here on account of the rapidity
variables already used in Eq.\ (\ref{10}).  In general, the valon
distribution of a hadron $h$ has a part specifying the
wave-function squared, $\tilde{G}_h$, and a part specifying
momentum conservation
\begin{eqnarray}
G_h(\xi _1,\cdots ) = \tilde{G}_h(\xi _1,\cdots)\ \delta\left(\sum_i
\xi _i -1 \right)  ,
\label{21}
\end{eqnarray}
where the functional form of $\tilde{G}_h$ is determined
phenomenologically.  Although for proton $\tilde{G}_p$ is found
to be highly nontrivial \cite{hy3}, for pion $\tilde{G}_{\pi}$
turns out to be very simple \cite{hy}
\begin{eqnarray}
\tilde{G}_{\pi} (\xi _1, \xi _2) = 1  ,
\label{22}
\end{eqnarray}
which is a reflection of the fact that the pion mass is much lower
than the constituent quark masses, so tight binding results in
large uncertainty in the momentum fractions of the valons.
Equation (\ref{22}) implies that the valon momenta of the pion is
uniformly distributed in the range $0 < \xi _i < 1$.

What remains in Eq.\ (\ref{19}) for us to determine is
$R_{\pi}^0$.  At this point we need to be more specific about the
quark and antiquark that recombine.  If the colors of $q$ and
$\bar{q}$ are considered, then the probability of forming a
color singlet pion is $1/9$ in $3 \times \bar{3}$.  Similarly, for
three quarks forming a proton the probability is $1/27$ in $3
\times 3
\times 3$.  In the parton distributions, $F_{q\bar{q}}$ for pion
production involves two color triplets and $F_{qqq}$ for proton
production involves three color triplets so the color factors work
out just right in that the factors of $9$ for $q\bar{q}$ and $27$
for $qqq$ are cancelled by the corresponding inverse factors in
the recombination probabilities.  In other words, for the $p/\pi$
ratio to be considered later, we can ignore the factors associated
with the color degrees of freedom and proceed with the
determination of $F_{q\bar{q}}$ without specifying the quark
colors and summing over them.

The situation with flavor is not the same.  For a  $u \bar{u}$ pair
and a $d\bar{d}$ pair, they can form $\pi^0$ and $\eta$ in the
flavor octet.  The branching ratio of $\eta$ to $3 \pi^0$ is 32.5\%
and to $\pi ^+ \pi ^- \pi ^0$ is 22.6\%.  Thus for every $\eta$
produced there is on average $1.2 \pi ^0$.  Due to the higher
mass of $\eta$ we make the approximation that the rate of
indirect production of $\pi ^0$ via $\eta$ is roughly the same as
the direct production from $u \bar{u}$ and $d \bar{d}$.  If we
now use $q\bar{q}$ to denote either  $u \bar{u}$ or $d
\bar{d}$, but not both $u \bar{u}$ and $d \bar{d}$,
then each pair of $q\bar{q}$ leads to one $\pi ^0$.  Since in a
heavy-ion collision there are many quarks and antiquarks
produced in the central region, it is reasonable to assume that the
$q$ distribution is independent of the $\bar{q}$ distribution so
that we can write $F_{q\bar{q}}$ in the factorizable form
\begin{eqnarray}
F_{q\bar{q}} \left(z _1, z _2\right) = F_q (z _1)\
F_{\bar{q}} ( z _2) ,
\label{23}
\end{eqnarray}
where $F_q$ stands for either $u$ or $d$ distributions, and
similarly for $F_{\bar{q}} $, but for $\pi ^0$
production $\bar{q}$ should be the antiquark partner of $q$.

The fact that we consider $\eta$ production above, but not the
vector meson $\rho$ requires an explanation. We defer that
discussion until  the next section, after we
have presented the formalism for the production of protons.

Returning now to the normalization of $R_{\pi}  \left(z _1, z
_2, z\right)$, we note that, using Eqs.\ (\ref{19}), (\ref{21}) and
(\ref{22}),
\begin{eqnarray}
\int dzzR_{\pi} (z _1, z_2, z) &=& \int {dz\over z} R_{\pi}^0
\delta \left({z _1 + z_2\over z}-1\right)\nonumber\\&& = R_{\pi}^0
\label{24}
\end{eqnarray}
is the probability that a $q$ at $z_1$ and a $\bar{q}$ at $z_2$
recombine to form a pion at any $z$.  According to our counting
in the second paragraph above, the total probability for $q\bar{q}
\rightarrow \pi ^0$ integrated over all momenta is
\begin{eqnarray}
\int^Z_0 {d z _1 \over  Z} \int^Z_0 {d z _2 \over  Z} \int dz\ z\,
R_{\pi}(z _1, z_2, z) = 1  ,
\label{25}
\end{eqnarray}
where $Z$ is the maximum $z_i$, whatever it is.  This
normalization condition is scale invariant, and we find, using
Eq.\ (\ref{24}), that
\begin{eqnarray}
R_{\pi}^0 = 1  .
\label{26}
\end{eqnarray}

Putting Eqs.\ (\ref{19}) - (\ref{23}) and (\ref{25}) in (\ref{15})
we obtain
\begin{eqnarray}
{dN_\pi  \over  zdz} = \int dz_1 dz_2{z_1z_2\over z}F_q(z_1)\,
F_{\bar{q}}(z_2)\,\delta(z_1 + z_2 - z).
\label{27}
\end{eqnarray}
This is obtained from Eq. (\ref{9}) where $y$ and $\phi$ are both
explicitly averaged over.  The LHS is to be identified with
$\Phi(z)$. Note that the $1/2\pi$ factors in Eqs.\ (\ref{1}) and
(\ref{3}), where $\Phi(z)$ is defined, are there to render $f(p_T,
s)$ an average distribution in $\phi$; that is the notation for the
experimental distribution, defined in \cite{ddl}. The distribution
defined by us in Eq.\ (\ref{9}) already includes the $1/2\pi$
factor, so our $dN_\pi/z\,dz$ is just the experimental $\Phi(z)$. As
we have mentioned earlier, the normalization of $z$ has no
significance.  By means of a scale change in Eq.\ (\ref{16}) we can
move from $z$ to
$x$, or vice-versa, without changing the scale invariant form of Eq.\
(\ref{27}).  In Eq.\ (\ref{6}) we found $\Phi (z)$ to have the form
\begin{eqnarray}
\Phi (z) = A \left(z^2 + c \right)^{-n} .
\label{28}
\end{eqnarray}
If we change $z$ to $x$ according to Eq.\ (\ref{16}), then by
keeping the total number of pions invariant, i.e.,
\begin{eqnarray}
\int dz\, z \,\Phi (z, K) = \int dx\, x \,\Phi^{\prime} (x,
K^{\prime})
 ,
\label{29}
\end{eqnarray}
we have
\begin{eqnarray}
\Phi^{\prime} (x, \lambda K) = \lambda^2\Phi (\lambda x, K).
\label{30}
\end{eqnarray}
It thus follows that
\begin{eqnarray}
\Phi^{\prime} (x) = \lambda^{2(1-n)} A \left(x^2 +
c/\lambda^2\right)^{-n}.
\label{31}
\end{eqnarray}
Similarly, in the $x$ variable the transformed quark distributions
is
\begin{eqnarray}
F^{\prime}_q (x_1, K^{\prime}) = \lambda^2  F_q (z_1, K).
\label{32}
\end{eqnarray}
Without having to specify the arbitrary scale factor $\lambda$,
let us work with the $z$ variable and rewrite Eq.\ (\ref{27}) as
\begin{eqnarray}
\Phi(z) = \int^z_0 dz_1\ z_1 \left(1 - { z_1 \over  z} \right)\ F_q
(z_1)\ F_{\bar{q}}(z - z_1)  .
\label{33}
\end{eqnarray}

We must now consider how the
$q$ and $\bar{q}$ distributions differ.  Unlike the structure
functions of the nucleon, where $q$ and $\bar{q}$ have widely
different distributions, we are here dealing with the partons at
high $p_T$ in heavy-ion collisions just before recombination.
The dynamics underlying their $p_T$ dependences is
complicated.  Many subprocesses are involved, which include
hard scattering, gluon radiation, jet quenching, gluon conversion
to quark pairs, thermalization, hydrodynamical expansion, to
name a few familiar ones.  At very large $p_T$ there are far more
quark jets than antiquark jets, since the valence quarks have
larger longitudinal momentum fractions than the sea quarks.  By
hard scattering the quarks therefore can acquire larger $p_T$
than the antiquarks.  Thus in that way one would expect the
$p_T$ distribution of the quarks to be very different from that of
the antiquarks.  However, that view does not apply to our
problem.  Those are the $q$ and $\bar{q}$ that initiate jets, along
with jets initiated by gluons.  The conventional approach is to
follow the jet production by jet fragmentation, which can
be modified by the dense matter that the initiating partons
traverse.  As discussed earlier, our approach is not to delve into
the dynamical origins of the $q$ and $\bar{q}$ distributions, but
to consider the recombination of $q$ and $\bar{q}$ just at the
point of hadronization.  Such $q$ and $\bar{q}$ are not the
partons that initiate jets, but are the parton remnants after the
hard partons radiate gluons which subsequently convert to
$q\bar{q}$ pairs.  Those parton remnants have similar
momentum distribution for $q$ and $\bar{q}$, since gluon
conversion creates $q$ and $\bar{q}$ on equal basis; those
partons are the ones that recombine to form hadrons.  They are
not to be confused with the jet-initiating hard partons that
fragment into hadrons in the fragmentation model.  In the
recombination picture those hard partons that acquire large $p_T$
immediately after hard scattering are not ready for
recombination; they lose momenta and virtuality through gluon
radiation until a large body of low-virtuality quarks and
antiquarks are assembled for recombination --- a view that is
complementary to the fragmentation picture.  Of course, there
are more quarks than antiquarks, since the number of valence
quarks of the participating nucleons cannot diminish.  For that
reason we allow $F_q(z)$ to differ in normalization from
$F_{\bar{q}}(z)$.  However, as a first approximation we assume
that their $z$ dependences are the same.

There is some indirect experimental evidence in support of our
assumption.  In Ref.\ \cite{ts} the $\bar{p}/p$ ratio for central
collisions is reported to be essentially constant within errors; more
precisely, it ranges between $0.6$ and $0.8$ for $p_T$ in the
range $0.5 < p_T < 3.8$ GeV/c.  Since $\bar{p}$ is formed by the
recombination of three $\bar{q}$, while $p$ is formed from three
$q$, a quick estimate of the $\bar{q}/q$ ratio is that it varies
between $0.6^{1/3}$ and $0.8^{1/3}$, i.e., from $0.843$ to
$0.928$.  Such a narrow range of variation is sufficient for us to
assume that $F_{\bar{q}}(z)$ has the same $z$ dependence as
$F_q(z)$.  For their relative normalization we take the mean
$\bar{p}/p$ ratio to be $0.7$.  Thus we adopt the $\bar{q}/q$
ratio to be
\begin{eqnarray}
F_{\bar{q}}(z)/F_q(z) = F^{\prime}_{\bar{q}}(x)/F^{\prime}_q(x)
=0.7^{1/3}  .
\label{34}
\end{eqnarray}
With this input we are finally ready to infer the quark
distribution from the pion distribution.
\begin{figure}[bth]
\includegraphics[width=0.55\textwidth]{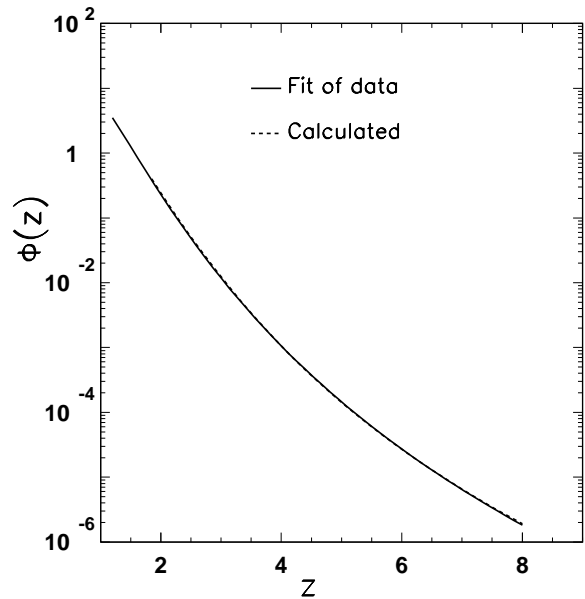}
\caption{The solid line is the fit of the data  as shown
in Fig.\ 1 (in a different scale), and the dashed line is the
theoretical calculation of $\Phi(z)$ using the quark distribution in
Fig.\ 4.}
\end{figure}

We parameterize $F_q(z)$ by
\begin{eqnarray}
F_q(z) = a \left(z^2 + z + z_0 \right)^{-m}
\label{35}
\end{eqnarray}
and adjust the three parameters $a$, $z_0$ and $m$ to fit
$\Phi (z)$ by using Eq.\ (\ref{33}).  We obtain an excellent fit
with the values
\begin{eqnarray}
a = 90 , \qquad z_0 = 1, \qquad m= 4.65  .
\label{36}
\end{eqnarray}
In Fig.\ 3 we show in solid line the data represented by the
formula in Eq.\ (\ref{6}) and in dashed line the result of the
theoretical calculation using Eqs.\ (\ref{33})-(\ref{36}).  They
coalesce nearly completely in the interval $1 < z < 8$.  The
quark distribution $F_q(z)$ is shown in Fig.\ 4.  To appreciate
the $p_T$ range corresponding to $z$ in Fig.\ 4, recall
 Eq.\ (\ref{2}), $p_T = zK$, and Fig.\ 2 for $K$.  Thus at
$\sqrt{s} = 200$ GeV, $p_T$ is $z$ in GeV.  Equations
(\ref{35}) and (\ref{36}) represent a main result of this study.
What is important is that we have found a scaling quark
distribution that is independent of $s$ from SPS to RHIC, and
perhaps to LHC.  It is a succinct summary of the effects of all the
dynamical subprocesses in heavy-ion collisions.  The non-trivial
$z$ dependence in Eq.\ (\ref{35}) indicates that there are
intrinsic scales in the low-$p_T$ problem.
\begin{figure}[tbh]
\includegraphics[width=0.55\textwidth]{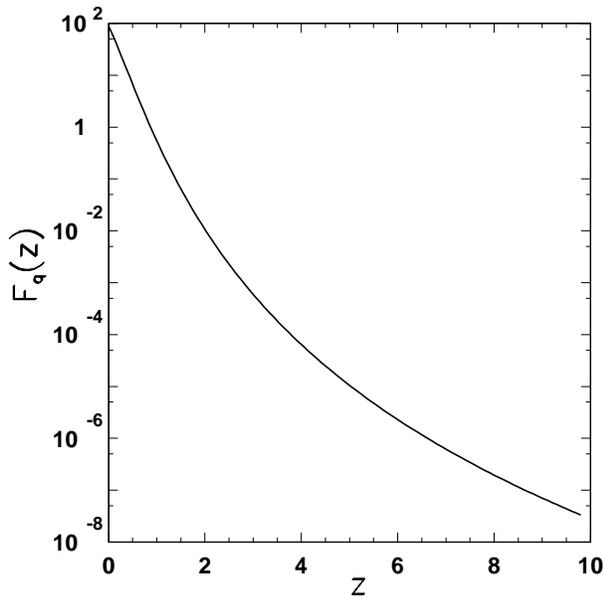}
\caption{Quark distribution in $z$.}
\end{figure}

\section{The $p/\pi$ Ratio}
The quark distribution obtained in the preceding section cannot
be checked directly.  Since it is the distribution at the end of its
evolution, massive dileptons would not be sensitive to it due to
their production at the early stages.  Proton production provides
the most appropriate test, since hadronization occurs near the
end.  We shall therefore calculate the proton distribution at high
$p_T$ and compare with the data on the $p/\pi$ ratio.  This is
not a completely satisfactory venture, since the proton mass is
large, so only at very high $p_T$ can our scale invariant
calculation be valid without explicit consideration of the mass
effect.  Present data on the $p/\pi$ ratio do not extend beyond
$p_T \sim 3.8$ GeV/c \cite{ts}.  Nevertheless, our calculation
should provide some sense on the magnitude of the rate of
proton production at the high $p_T$ end.

The inclusive distributions in the scaled $p_T$ variable can be
obtained in the recombination model by generalizing Eq.\
(\ref{13}) to the recombination of three quarks
\begin{eqnarray}
{dN_p \over zdz} &=& \int dz_1 dz_2 dz_3\ z_1\,z_2\,z_3\nonumber\\
&& F(z_1, z_2, z_3)\ R_p(z_1, z_2, z_3, z)
\label{37}
\end{eqnarray}
where $F(z_1, z_2, z_3)$ is given the factorizable form
\begin{eqnarray}
F(z_1, z_2, z_3) = F_u (z_1) F_u (z_2) F_d (z_3).
\label{38}
\end{eqnarray}
As in Eq.\ (\ref{19}) we relate the recombination function $R_p$
to the valon distribution, $G_p$, of the proton
\begin{eqnarray}
R_p(z_1, z_2, z_3, z) = R_p^0 \ z^{-2}\,G_p(\xi_1,\xi_2, \xi_3)  ,
\label{39}
\end{eqnarray}
where $G_p$ has the general form given in Eq.\ (\ref{21}), and
$R_p^0$ remains to be determined.  In Ref.\ \cite{hy3} a detailed
study of the proton structure functions has been carried out in
deriving the valon distribution from the parton distributions that
fit the deep inelastic scattering data.  It is
\begin{eqnarray}
\tilde{G}_p (\xi_1, \xi_2, \xi_3) = g  \ (\xi_1\,
\xi_2 ) ^\alpha\ \xi_3^{\beta}  ,
\label{40}
\end{eqnarray}
where
\begin{eqnarray}
\alpha = 1.755,  \qquad \beta = 1.05  ,
\label{41}
\end{eqnarray}
\begin{eqnarray}
g = \left[B \left(\alpha + 1, \beta +1 \right) B \left(
\alpha + 1, \alpha + \beta +2\right)\right]^{-1}  .
\label{42}
\end{eqnarray}
Single-valon distributions $G_p(\xi_i)$ can be
obtained from the three-valon distribution by integration and are
peaked around $\xi = 1/3$, indicating that each of the three valons
carries on average roughly $1/3$ the momentum of the proton, their
sum being strictly 1.  Details of the valon model, described in
\cite{hy3}, are not needed for the following.  It is only necessary
to recognize that the recombination of two $u$ quarks with a
$d$ quark to form a proton has a probability proportional to
the proton's valon distribution that accounts for the proton
structure.  The other point to bear in mind is that the valon
distribution in the proton is obtained in the frame where the
proton momentum is infinitely large so the finite masses of the
proton and valons are unimportant.  However, the validity of
that result when the proton momentum is only two or three times
larger than its mass is questionable.  With that caveat we proceed
with our scale invariant calculation and see what can emerge.

As discussed in the preceding section, there is no need to consider
the color factors for either pion or proton formation since
hadrons are color singlets, but the flavor octets for these hadrons
do introduce some factors.  The $\left.|uud\right>$ state
appears in
$10 + 8 + 8 ^{\prime}$ of $3 \times 3 \times 3$; among them
the first two contain $\Delta ^+$ and $p$.  Thus the flavor parts
of
$|\left<\Delta ^+\left|uud\right>|^2\right.$ and
$|\left<p \left| uud\right>|^2\right.$ are $1/3$ for
each.  Since
$\Delta ^+$ decays to $p + \pi^0$ and $n + \pi^+$,
$|\left<p\left| \Delta ^+\right>|^2\right.$ gives another
factor
$1/2$.  The spin decomposition of $2 \times 2 \times 2$ is $4
+ 2
+ 2$, among which the $\Delta ^+$ component is $4/8$ and
$p$ is $2/8$.  Putting the flavor and spin factors together, we
have
\begin{eqnarray}
&&\left| \left<p \left| uud \right. \right>\right|^2 +
\left| \left<p\, | \Delta ^+\right>\left< \Delta
^+ \left| uud \right. \right>\right|^2 \nonumber\\
&&\, \, \, \, \,= {1  \over  3}\times {1
\over  4} + {1  \over  3}\times {1 \over  2} \times {1 \over  2} =
{1 \over  6}  .
\label{43}
\end{eqnarray}
We thus normalize $R_p$, as we have done in Eq.\ (\ref{25}), by
\begin{eqnarray}
\int^Z_0 \prod^3_{i = 1} {dz_i  \over  Z} \int dz\, z \,R_p
(z_1, z_2, z_3, z) = {1 \over  6} .
\label{44}
\end{eqnarray}
In view of Eqs.\ (\ref{21}) and (\ref{39}) we have
\begin{eqnarray}
R_p^0\int^Z_0 \prod^3_{i = 1} {dz_i  \over  Z}\ \tilde{G}_p
\left({z_1 \over z_t}, {z_2\over z_t}, {z_3\over z_t}\right) = {1
\over  6},
\label{45}
\end{eqnarray}
where $z_t = \sum_i z_i$.  Using Eq.\ (\ref{40}), the above
integral can be transformed to
\begin{eqnarray}
g \int^1_0 \prod^3_{i = 1} d \zeta _i \left({ \zeta_1 \zeta_2
\over  \zeta ^2_t} \right)^{\alpha} \left({ \zeta_3
\over  \zeta _t} \right)^{\beta} = 2.924
\label{46}
\end{eqnarray}
with $\zeta _i = z_i/Z$ and $\zeta _t = \sum _i \zeta _i$.  There
is no explicit dependence on $Z$, and Eqs.\ (\ref{41}) and
(\ref{42}) have been used in getting the numerical value in Eq.\
(\ref{46}).  It thus follows that
\begin{eqnarray}
R_p^0 = 0.057  .
\label{47}
\end{eqnarray}

At this point we should address the question why we consider
$\Delta^+$ production above, but not $\rho$ production in the
preceding section. For the production of $\pi^0$, if we are
to consider the contribution from $\rho^\pm$ (since $\rho^0$ does
not decay strongly into $\pi^0$), we would be extending our scope to
other flavored states besides $u\bar u$ and $d\bar d$. Then other
vector mesons and higher resonances, such as $K^*$, that can decay
into
$\pi^0$ must also be included.  Similarly, for $p$ production the
consideration of other states beside $uud$ would involve many
resonances that can decay into $p$. The system is not closed
without more phenomenological input beside $\pi^0$. Thus for a
closed system in which a prediction can be made, we limit ourselves
to only the $u\bar u$ and $d\bar d$ in the meson states and $uud$
in the baryon states; hence, only $\pi^0, \eta, p$ and $\Delta^+$
are considered. To include $u\bar d$ and $d\bar u$, we must also
include $uuu$ and $udd$, and so on.   We surmise that if more
resonances are included in both the meson and baryon sectors, the
$p/\pi$ ratio to be determined below would change somewhat; however,
the result is not likely to differ by a factor greater than 2.

With the recombination function $R_p$ completely determined,
and the quark distribution $F_q \left(z_i\right)$ given by Eqs.\
(\ref{35}) and (\ref{36}), we can now use Eq.\ (\ref{37}) to
calculate the proton distribution in $z$.  The result is shown by
the solid line in Fig.\ 5, where only the portion $z > 2$ is
exhibited.  We have stated at the outset that the scale invariant
form of $dN_p/zdz$ cannot be expected to be valid when the
mass effect is important.  The relevant value of $z$ corresponding
to the proton mass (let alone the $\Delta ^+$ mass) is
\begin{eqnarray}
z_m = m_p/K  ,
\label{48}
\end{eqnarray}
which ranges from 1.3 at $\sqrt{s} = 17$ GeV to 0.94 at 200 GeV.
As expected, the scaling violating effects are energy dependent.
Thus we should not regard the calculated result to be reliable for
$z < 3$.  At very low $p_T$ the distributions of all hadrons can
be given exponential fits in the transverse mass.  The STAR data
for most central collisions at $\sqrt{s}= 130$ GeV \cite{pj} give
for $\bar{p}$ production for $p_T < 0.6$ GeV
\begin{eqnarray}
{1  \over  2 \pi m_T}{d^2N_{\bar{p}}  \over  dm_T dy} = 4 \exp
\left[-\left(m_T-m_p\right)/T_p\right]
\label{49}
\end{eqnarray}
where $m_T = \left(m^2_p  + p^2_T\right)^{1/2}$ and $T_p =
565$ MeV.   To convert this distribution to that for $p$ we assume
that only the normalization at $p_T = 0$ needs to be adjusted.
The $\bar{p}/p$ ratio at low $p_T$ is 0.6 \cite{ts}.  Since
$m_Tdm_T = p_T dp_T$ and the distribution in $p_T$ changes
by a scale factor $K^2$ given in Eq.\ (\ref{2}), where $K = 0.9$
for $\sqrt{s} = 130$ GeV, the factor 4 in Eq.\ (\ref{49}) should
therefore be changed to $4 \times 0.81/0.6 = 5.4$.   Expressing
$m_T$ in terms of $z$ by use of Eq.\ (\ref{2}) with $K = 0.9$,
we show the $z$ dependence of the distribution for $p$ in Fig.\ 5
by the short dashed line.  The region $0.5 < z <2$ is left
blank because our scaling result cannot be reliably extended into
that region.  Nevertheless, it is gratifying to observe that the
theoretical calculation without any free parameters produces a
proton distribution at large $z$ that is reasonable in
normalization and shape and can smoothly be connected with the
low-$p_T$ distribution by interpolation.
\begin{figure}[htb]
\includegraphics[width=0.55\textwidth]{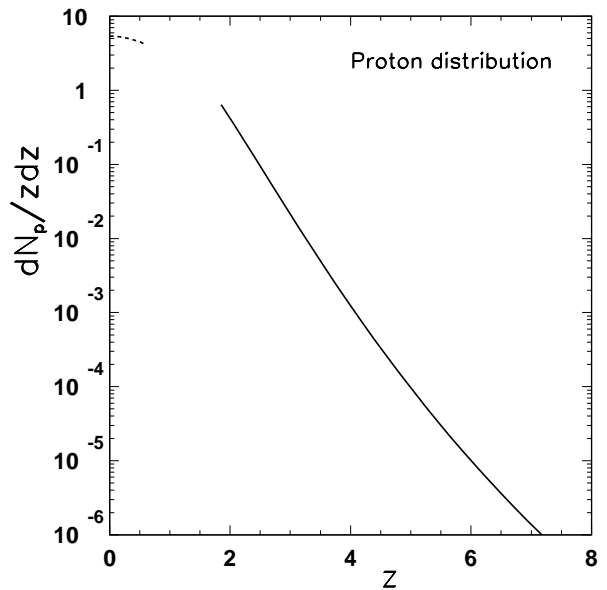}
\caption{Proton distribution in $z$. Solid line is the
theoretical result; the dashed line is the fit of  data at low-$p_T$
\cite{pj}.}
\end{figure}

With the proton distribution now at hand, we can calculate the
$p/\pi$ ratio.  For the pion distribution we use $\Phi (z)$ given
in Eq.\ (\ref{6}).  For proton we use the calculated result based
on Eq.\ (\ref{37}).  Their ratio, defined by
\begin{eqnarray}
R_{p/\pi}(z) = {dN_p  \over  zdz} / \Phi (z)
\label{50}
\end{eqnarray}
is shown by the solid line in Fig.\ 6.  The preliminary data on the
$p/\pi$ ratio were reported in Ref.\ \cite{ts}, which we show also
in Fig.\ 6 for both $\sqrt{s} = 130$ and $200$ GeV.  Note that
because it is a ratio there is no change in the normalizations of
$R_{p/\pi}$ for the two energies, but in transforming from
$p_T$ to $z$ the factor
$K$ in Eq.\ (\ref{2}) must be taken into account.  Unlike the
pion case the effects of the proton mass are not negligible for
$p_T
\stackrel{<}{\sim} 3$ GeV/c, and one sees no scaling in $s$ or
$z$ in Fig.\ 6.  Our scale invariant calculation is unreliable for
$z < 3$ and shows a result that is obviously too high at $z
\stackrel{<}{\sim} 2$.  There seems to be a good chance that
the theory and experiment can agree well for $z > 4$.  In Fig.\ 6
we show two curves that can connect our scaling result with the
data.  The dotted curve is an eyeball fit of the 130 GeV data with
a connection at $z= 3.5$, while the dashed curve fits the 200 GeV
data with a connection at $z = 4$.  In the absence of a
theoretical study that takes the mass-dependent effects into
account in the intermediate $p_T$ region, the only point we
can make here is that it is not hard to produce a
$p/\pi$ ratio that exceeds 1 in the scale invariant calculation in
the recombination model, but it does so in a region where both
theory and experiment need refinement.  Judging by what is
self-evident in Fig.\ 6, we see no strong need for any exotic
mechanism for proton production
(as proposed in \cite{vg}) beyond the conventional
subprocess where three quarks recombine to form a proton.
\begin{figure}[bth]
\includegraphics[width=0.55\textwidth]{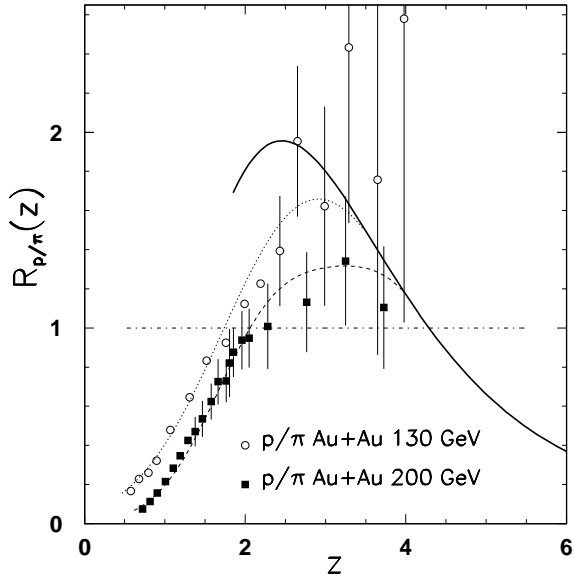}
\caption{Proton-to-pion ratio: solid line is the scaling
distribution from calculation; data (preliminary) are from Ref.\
\cite{ts}. The dotted and dashed lines are eyeball fits of the data
as extrapolations from the scaling result.}
\end{figure}

\section{Conclusion}

The discovery of a scale invariant distribution $\Phi (z)$ for pion
production at intermediate and high values of $p_T$ in
heavy-ion collisions ranging over energies in excess of an order of
magnitude of variation is an important phenomenology
observation that should be checked experimentally in great
detail.  Additional energy points should be added not only to
strengthen the validity of the scaling behavior, but also
to find the onset of scaling violation, if it exists.

The phenomenological properties of hadron
production provide useful insights into the hadronization process
and  into the nature of the quark system just before they turn
into hadrons. The usual approach to the study of heavy-ion
collisions is from inside out, following the evolution of the dense
matter, either in terms of hydrodynamical flow or of hard parton
scattering and subsequent hadronization by fragmentation \cite{gvw}.
Our approach pursued here is from outside in, by starting from the
observed scaling behavior of the pions produced and deriving the
momentum distribution of the quarks that can give rise to such a
behavior.  That is accomplished by use of the recombination
model.  There is no direct way to check the validity of the quark
distribution thus obtained.  However, we have used it to determine
the proton distribution at high $p_T$ where the mass effects are
unimportant.  The data on proton production have not yet
reached that regime where the predicted scaling distribution can
be checked.  In the region where data exist on the $p/\pi$ ratio
we find that our calculated result, though not reliable, is in rough
agreement with the imprecise data to the extent that the ratio
exceeds 1, a feature  that is notable.

While the recombination model needs further work to take the
proton mass into account at intermediate $p_T$, its formulation
in the invariant form has been developed here to treat the very
high $p_T$ region.  We have made the assumption that the quark
and antiquark distributions are the same, apart from
normalization, just before recombination.  That assumption is
supported by the constancy of the $\bar{p}/p$ ratio in the
PHENIX data in the central region.  That experimental fact can
also be used to lend credence to our general approach to
hadronization that is treated as a recombination process, for
which we have given arguments why the distributions of quarks
and antiquarks should be similar before they recombine.  In
contrast, the fragmentation model would suggest a decreasing
function of $\bar{p}/p$ in $p_T$ because of the dominance of
quark jets over antiquark jets at large $p_T$ \cite{jet}.

In this paper we have only considered the energy dependence of the
$p_T$ spectrum at fixed maximum centrality. It is natural to ask
what the dependence is on centrality. We have investigated that
problem by making a phenomenological analysis of the data on
centrality dependence without using any hadronization model, and
found a scaling behavior very similar to what is reported here.
The scaling distribution found there \cite{hy4}
includes the very small $p_T$ region in the fit, and is therefore
more accurate. But the fits in the intermediate and large $p_T$
region are the same. The implication on the centrality dependence
of the $p/\pi$ ratio in the recombination model is still under study.

To have an invariant quark distribution independent of $s$ just
before hadronization provides an unexpected picture of the
quark system.  It suggests that the evolution of the system
proceeds toward a universal form whatever the collision energy
may be.  We expect that universal form to depend on rapidity.
The origin of such a scaling distribution in $z$ is not known at this
point and can form the focus of a program of  future theoretical
investigations.

\section*{Acknowledgment} We wish to thank D.\ d'Enterria
for a helpful communication.  This work was supported, in part,
by the U.\ S.\ Department of Energy under Grant No.
DE-FG03-96ER40972.

\end{document}